# Nodeless *d*-wave pairing in monolayer FeSe revealed by intra-unit-cell Josephson interferometry

**Authors:** Runxin Li[1*], Yuchen Huo[2,1*], Jiong Mei[3*], Ming-Qiang Ren[1*†], Miaomiao Wang[1], Quanxing Wei[1,4], Fengyu Yao[1], Maowen Zheng[1,4], Qing Qu[1], Cui Ding[4], Hongtao Yuan[5,6], Zhuoyu Chen[1,4], Wei-Qiang Chen[1,4], Kun Jiang[3,7†], Qi-Kun Xue[1,4,8,9] and Yu-Jie Sun[1,4,10†]

**Affiliations:**

[1]State Key Laboratory of Quantum Functional Materials, Department of Physics, and Guangdong Basic Research Center of Excellence for Quantum Science, Southern University of Science and Technology, Shenzhen 518055, China.

[2]Department of Physics, Tsinghua University, Beijing 100084, China.

[3]Beijing National Laboratory for Condensed Matter Physics, and Institute of Physics, Chinese Academy of Sciences, Beijing 100190, China.

[4]Quantum Science Center of Guangdong-Hong Kong-Macao Greater Bay Area, Shenzhen 518055, China.

[5]National Laboratory of Solid State Microstructures, College of Engineering and Applied Sciences, Nanjing University, Nanjing 210000, China.

[6]Jiangsu Physical Science Research Center, Nanjing University, Nanjing 210000, China.

[7]School of Physical Sciences, University of Chinese Academy of Sciences, Beijing 100190, China.

[8]Frontier Science Center for Quantum Information, Beijing 100084, China.

[9]Beijing Academy of Quantum Information Sciences, Beijing 100193, China.

[10]Center for Advanced Light Source, Southern University of Science and Technology, Shenzhen 518055, China.

[*]These authors contributed equally to this work.

[†]Corresponding authors. E-mails: renmq@sustech.edu.cn; jiangkun@iphy.ac.cn; sunyj@sustech.edu.cn

**The Josephson effect provides a direct phase-sensitive probe of superconducting order and has played a central role in identifying unconventional pairing symmetries[1-6]. Conventional Josephson interferometry, however, extracts phase information through engineered junction geometries and crystallographic orientations, rather than by resolving the internal phase structure of the pair state within a single unit cell[2-6]. Here, we use a superconducting Nb tip to realize atomic-scale Josephson interferometry in monolayer FeSe/$SrTiO_3$. Despite a fully gapped**

**quasiparticle spectrum, the local Josephson response exhibits a striking intra-unit-cell pattern, with maxima centred on Fe–Fe bonds and robust nodal lines along the Fe–Se directions. We show that these nodal lines arise from destructive interference between Josephson tunnelling amplitudes associated with opposite-sign pairing components. Microscopic calculations reproduce the observed spatial pattern and identify it as a fingerprint of nodeless *d*-wave pairing. Our results establish intra-unit-cell Josephson interferometry as a local phase-sensitive probe of pairing symmetry in unconventional superconductors.**

The relative phase between microscopic pairing components is a defining property of an unconventional superconducting state and provides a direct fingerprint of its pairing symmetry[7,8]. Traditional quasiparticle probes such as angle-resolved photoemission spectroscopy (ARPES) and scanning tunnelling microscopy/spectroscopy (STM/S) resolve the momentum- and real-space structure of the superconducting excitation gap but are primarily sensitive to its magnitude[9,10]. These probes, therefore, cannot distinguish pairing states with similar gap magnitudes but different relative phases between their constituent components. The Josephson effect provides a fundamentally different probe because Cooper-pair tunnelling is intrinsically phase sensitive[1]. Indeed, corner-junction, superconducting quantum interference, and tricrystal experiments were instrumental in establishing the sign-changing *d*-wave order parameter of the cuprates[2-6]. However, these approaches extract phase information through carefully engineered junction geometries and crystallographic orientations, making them technically demanding and sensitive to junction details. More fundamentally, they probe phase interference averaged over mesoscopic or macroscopic junction regions and therefore cannot directly resolve the microscopic phase structure of the pair state.

Superconducting-tip STM instead combines the phase sensitivity of Josephson tunnelling with atomic-scale spatial resolution, enabling local measurements of coherent Cooper-pair transport[11-15]. Previous studies have used this capability to image pair-density-wave order[16,17], resolve atomic-scale variations in the local Josephson response[18-20], and probe frustrated coupling between multiple superconducting condensates[21]. Here, we extend Josephson imaging to the intra-unit-cell scale, enabling direct access to the microscopic pairing structure of the superconducting state. The local Josephson amplitude reflects the coherent interference of contributions from distinct pairing components, effectively forming a local Josephson interferometer. The relative contributions of these components evolve with tip position, thereby mapping their internal phase structure into a spatially resolved Josephson interference pattern.

Monolayer FeSe on $SrTiO_3$ (STO), an interfacial high-temperature superconductor[22],

provides a stringent test of this approach: its superconducting state is fully gapped, yet the underlying pairing symmetry remains unresolved[23-36]. Its low-energy electronic structure is dominated by electron pockets near the Brillouin-zone (BZ) boundaries[23-26]. For this Fermi-surface geometry, pairing states with distinct internal sign structures can produce similarly nodeless excitation spectra[32,33]. In particular, a $d$-wave state can remain gapped because its symmetry-imposed nodal lines do not intersect these electron pockets[37-40]. Gap magnitude measurements alone cannot distinguish between a sign-preserving state and a nodeless sign-changing state. Access to the relative phase of the pair state is therefore essential for identifying the pairing symmetry in monolayer FeSe.

Here we use a conventional $s$-wave superconducting Nb tip to realize intra-unit-cell Josephson interferometry in monolayer FeSe on STO. The tip forms a movable vacuum Josephson junction with the sample (Fig. 1a), allowing us to map the pair-tunnelling response continuously across the FeSe lattice. We observe a pronounced intra-unit-cell pattern, with maxima at nearest-neighbour Fe–Fe bond centres and robust nodal lines along the Fe–Se directions. Microscopic tunnelling calculations reproduce this spatial structure for a nodeless $d$-wave state, and show that the nodal lines arise from destructive interference between tunnelling amplitudes associated with opposite-phase pairing components. Our results demonstrate that intra-unit-cell Josephson interferometry can directly access the internal phase structure of an unconventional superconducting state in real space.

**Establishing Josephson tunnelling in monolayer FeSe**

Atomically resolved topography of monolayer FeSe grown on STO substrates reveals the characteristic Se-terminated surface lattice (Fig. 1b), providing a crystallographic reference for the spatially resolved measurements below. Tunnelling spectra acquired with a normal-metal tip reveal a fully gapped superconducting state, with characteristic gap energies of $\sim 10$ and $\sim 15$ meV (Supplementary Fig. 1), consistent with previous STM measurements[22,27,32,41]. The superconducting Nb tip was characterized independently on a Nb single crystal and exhibits a gap of $\Delta_{tip} \sim 1.4$ meV (Supplementary Fig. 2). The resulting Nb-tip–FeSe junction enables Cooper-pair tunnelling while retaining the atomic spatial resolution of STM.

Differential-conductance spectra, $g(V) \equiv dI/dV$, acquired at a fixed position, reveal two distinct low-energy features (Fig. 1c). A pronounced zero-bias peak is flanked by symmetric negative-differential-conductance dips at ~ ±0.4 meV, characteristic of Josephson tunnelling[13,17,19-21]. We denote the zero-bias and dip conductances by $g_0$ and $g_{dip}$, respectively. The corresponding $I{-}V$ curves exhibit a double-kink structure, featuring a pair of current extrema associated with the Josephson response, further supporting this assignment (Supplementary Fig. 3). Additional conductance peaks, denoted $g_A$, appear near $eV = \pm\Delta_{tip}$, consistent with Andreev processes[42]. Further quantitative analysis reveals that both $g_0$ and $g_A$ scale linearly with $1/R_N^2$ (Fig. 1d). This behaviour is consistent with phase-diffusive Josephson tunnelling and the lowest-order Andreev process, respectively[43,44]. Although microscopically distinct, both processes involve Cooper-pair transfer and therefore exhibit the observed quadratic dependence on junction conductance.

**Imaging the intra-unit-cell Josephson response**

Having established $g_0$ as a local measure of Josephson pair tunnelling, we next image the Josephson response within individual FeSe unit cells (Fig. 2a,b). Strikingly, despite the fully gapped quasiparticle spectrum of monolayer FeSe, the real-space Josephson response develops pronounced intra-unit-cell modulation with distinct nodal lines, pointing to an internal phase structure of the pair state rather than a vanishing gap magnitude. Using the well-established surface registry[22,32,41,45] together with the simultaneously acquired atomic-resolution topography (Fig. 2a), we find that these nodal lines run along the Fe–Se directions (Fig. 2b), enclosing four response maxima centred near the nearest-neighbour Fe–Fe bond.

Site-dependent tunnelling spectra (Fig. 2e) directly resolve the spectroscopic contrast across the intra-unit-cell Josephson modulation. At the Fe–Fe bond-centre positions, 1–3, as indicated in the inset, all spectra display the characteristic low-energy features identified in Fig. 1c: a pronounced zero-bias Josephson peak $g_0$, the accompanying negative-differential-conductance $g_{dip}$, and finite-bias Andreev features $g_A$. By contrast, the Josephson peak and dips are nearly absent at the Fe and Se sites, where the Andreev features are also substantially reduced. These spectra therefore confirm that the nodal lines in $g_0(\boldsymbol{r})$ correspond to a genuine suppression of the local Josephson response.

Two representative line profiles extracted from $g_0(\boldsymbol{r})$ map further resolve the nodal structure (Fig. 2f). Along a trajectory passing through Fe–Fe bond centres (cyan-blue line), $g_0$ exhibits a periodic sequence of pronounced maxima and zero minima. By contrast, along the Fe–Se direction (orange line), the Josephson response is strongly suppressed and nearly zero. These profiles directly establish that the nodal lines of the local Josephson response extend along the Fe–Se directions. We note that the four maxima at the Fe–Fe bond centres show some variation in relative intensity between measurements, which may reflect either tip-dependent tunnelling details or intrinsic symmetry breaking, while the nodal-line positions and symmetry remain robust (Supplementary Fig. 4).

The negative-differential-conductance response $g_{dip}(\boldsymbol{r})$, acquired over the same field of view (Fig. 2c), exhibits approximately reversed contrast relative to $g_0(\boldsymbol{r})$, rendering the nodal lines particularly prominent as bright streaks along the Fe–Se directions. The finite-bias Andreev conductance $g_\mathrm{A}(\boldsymbol{r})$ likewise displays a pronounced intra-unit-cell modulation (Fig. 2d), with enhanced response near Fe–Fe bonds and suppression along Fe–Se directions. The Josephson and Andreev responses thus exhibit similar spatial patterns within the unit cell.

To determine whether the spatial nodal structure is an inherent feature of intra-unit-cell Josephson tunnelling, we perform the same measurement on the conventional $s$-wave superconductor $NbSe_2$. The simultaneously acquired topography resolves the atomic lattice (Fig. 2g), whereas the zero-bias Josephson conductance $g_0(\boldsymbol{r})$ shows only weak intra-unit-cell variation (Fig. 2h). The corresponding line profiles remain finite with no nodal minima, in sharp contrast to monolayer FeSe (Fig. 2i). This result rules out a generic modulation imposed by the superconducting tip, atomic corrugation or the vacuum tunnelling geometry and identifies the nodal Josephson pattern as an intrinsic property of monolayer FeSe.

**Origin of the intra-unit-cell Josephson response**

These measurements establish two key properties of superconductivity in monolayer FeSe. First, tunnelling spectroscopy reveals a fully gapped quasiparticle spectrum. Second, the intra-unit-cell Josephson response exhibits robust nodal lines along the Fe–Se directions (Fig. 3a). Experimentally, the emergence of such nodal lines points to destructive interference between microscopic pairing components carrying relative phase differences. The observations therefore reveal a superconducting state that combines a nodeless excitation spectrum with a non-trivial internal phase structure of the order parameter.

Monolayer FeSe provides a particularly relevant setting in which this constraint can be realized. Within the unfolded one-Fe BZ convention[33-40], the low-energy electronic structure is dominated by electron pockets centred at the X and Y points, whereas the hole bands around Γ lie well below the Fermi level (Fig. 3b)[23-25]. For this unusual Fermi-surface topology, the symmetry-imposed nodal lines of a $d$-wave order parameter $\Delta(\boldsymbol{k}) \propto \cos k_x - \cos k_y$ (dashed lines), avoid these electron pockets. The quasiparticle spectrum can therefore remain fully gapped even though the order parameter changes sign between the two electron-pocket sectors. The nodal directions of the $d$-wave form factor correspond to the Fe–Se directions identified experimentally in the Josephson response.

To test whether this pairing state accounts for the observed real-space structure, we calculate the local Josephson response using a minimal model of monolayer FeSe[40].

The anomalous pair correlations of the superconducting state are projected into the vacuum through FeSe Wannier functions and coupled to the Nb tip[12-15]. The full microscopic formulation is provided in the Methods and Supplementary Information Sections S1–S4.

The calculated normalized Josephson critical-current map, $|I_J(\boldsymbol{r})|/|I_J|_{\max}$, for the *d*-wave pairing state is shown in Fig. 3c. It closely reproduces the principal features of the experimental map in Fig. 3a: the response is strongest near the Fe–Fe bond centres and is suppressed to zero along the Fe–Se directions. The line profiles in Fig. 3d further illustrate this correspondence. Along a trajectory through successive Fe–Fe bond centres (cyan-blue line), the calculated response exhibits a periodic sequence of pronounced maxima, resembling the experimental $g_0(\boldsymbol{r})$ profile in Fig. 2f. By contrast, along the Fe–Se direction (orange line), the calculated Josephson response remains zero, reproducing the experimentally observed nodal direction.

As a reference test of the tunnelling framework, we perform the same approach for conventional *s*-wave $NbSe_2$. The calculated response exhibits only weak atomic-scale modulation and no symmetry-related nodal structure (Fig. 3e,f), consistent with the $NbSe_2$ measurements in Fig. 2g–i. The agreement between experiment and calculation in both superconductors supports the interpretation of the FeSe nodal pattern as interference between opposite-sign pairing components, rather than a generic feature of intra-unit-cell Josephson imaging.

**Real-space Josephson interference**

The real-space structure of the *d*-wave pairing state directly explains the intra-unit-cell Josephson pattern. Fourier transforming the $\cos k_x - \cos k_y$ form factor gives nearest-neighbour Fe–Fe bond pairing components with opposite phases, represented by $+\Delta$ (red bonds) and $-\Delta$ (blue bonds) shown in Fig. 4a. A superconducting tip at position $\boldsymbol{r}$ generally couples to several nearby bond-centred pairing components, so the local Josephson amplitude is a coherent sum of contributions with different phases. Schematically, $I_J(\boldsymbol{r}) \sim (w_+(\boldsymbol{r}) - w_-(\boldsymbol{r}))\Delta$, where $w_+(\boldsymbol{r})$ and $w_-(\boldsymbol{r})$ denote the position-dependent tunnelling weights for the two opposite-phase components.

This simple picture explains the observed spatial pattern. Above an Fe site (Fig. 4b), the tip couples with approximately equal weights to opposite-phase pairing components, leading to destructive interference and a strongly suppressed Josephson response. As the tip moves towards the centre of an Fe–Fe bond (Fig. 4c), the two tunnelling weights become unequal, lifting the cancellation and producing a pronounced Josephson maximum. Above a Se site (Fig. 4d), the symmetry of the idealized junction restores equal weights and hence destructive interference. The bond-centred maxima and intervening nodal lines thus reflect the changing balance between finite, opposite-phase pair-tunnelling contributions as the tip moves within the unit cell. In this picture, a Josephson node marks cancellation of the net tunnelling amplitude without requiring the local superconducting gap to close.

Mirror symmetry further explains why this cancellation extends along well-defined crystallographic lines. The relevant mirror operations leave the Fe–Se directions invariant while exchanging the two opposite-phase bond components. Along the Fe–Se lines, their tunnelling weights are therefore constrained to be equal, $w_+(\boldsymbol{r}) = w_-(\boldsymbol{r})$, and the corresponding Josephson amplitudes cancel. Away from these mirror lines, the two components are no longer sampled equally, $w_+(\boldsymbol{r}) \neq w_-(\boldsymbol{r})$, and a finite Josephson coupling develops. The nodal lines therefore arise from opposite pairing phases combined with mirror-enforced equal tunnelling weights, rather than from accidental minima in the tunnelling matrix element.

This symmetry requirement distinguishes the $d$-wave state from the principal $s$-wave pairing candidates proposed for monolayer FeSe. A conventional sign-preserving $s$ state, a bonding–antibonding $s_{\pm}$ state and an incipient $s_{\pm}$ state all transform trivially[32,46-49]. Although the latter two contain sign reversals between different electronic sectors, those sign changes do not generate the same mirror-related cancellation. Corresponding calculations (Supplementary Fig. 5) produce intra-unit-cell modulations but no nodal lines along the experimentally observed Fe–Se directions. The measurement therefore probes not simply whether the superconducting order parameter changes sign, but also how its sign structure is organized within the microscopic pair state.

Our results establish intra-unit-cell Josephson interferometry as a local phase-sensitive probe of the microscopic phase structure underlying superconducting pairing symmetry. Applied to monolayer FeSe, this approach reveals the relative phase between microscopic pairing components and identifies a nodeless $d$-wave superconducting state. More broadly, it provides a clean phase-sensitive route to resolving competing pairing symmetries in unconventional superconductors, including candidate $p$-wave and twist-engineered systems.

## Figure legends

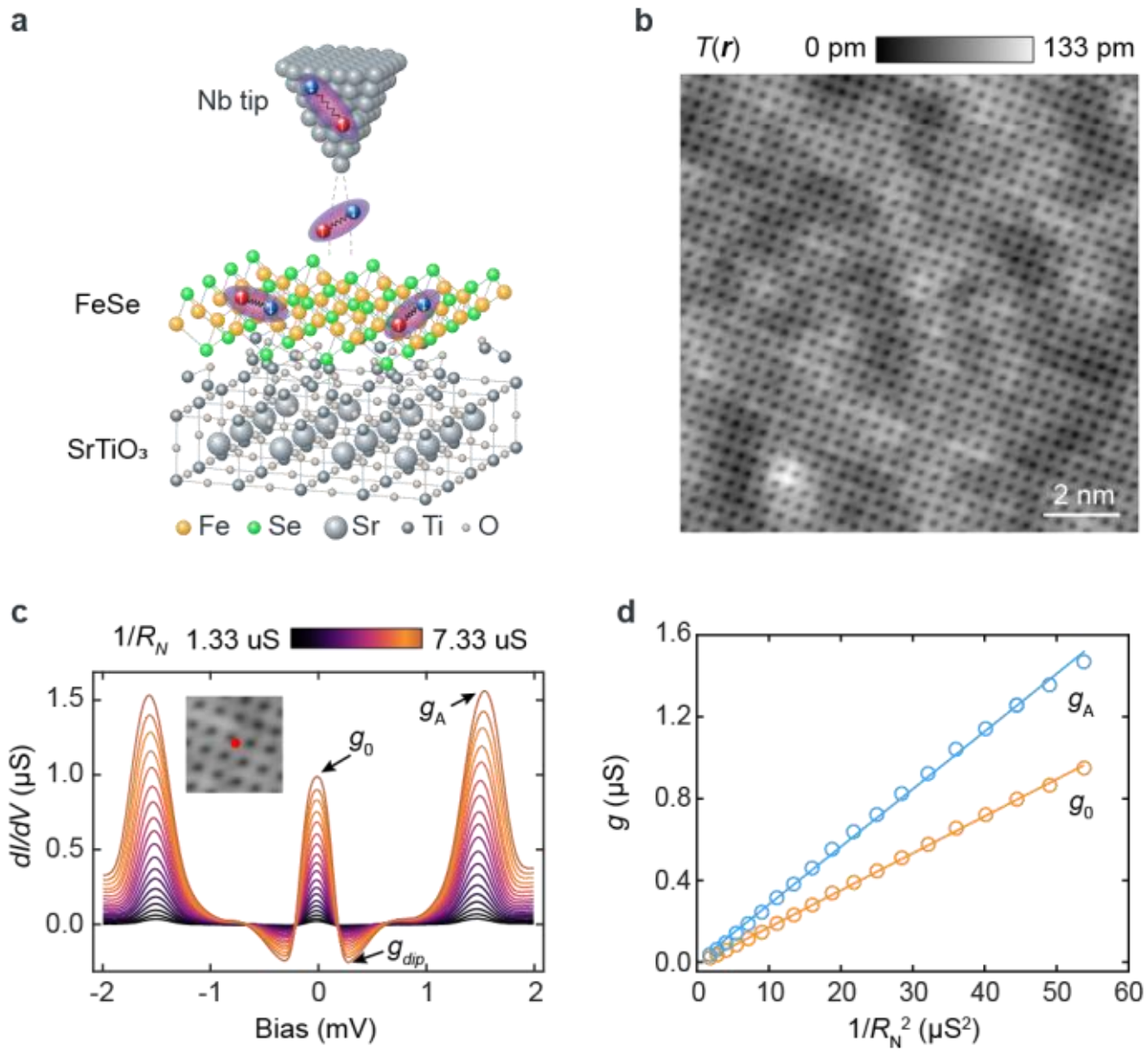

**Fig. 1 | Josephson tunnelling in monolayer FeSe/STO. a,** Schematic of the Josephson STM experiment, in which a superconducting Nb tip forms an atomically localized vacuum junction with monolayer FeSe grown on STO. **b,** Atomically resolved topography, $T(\boldsymbol{r})$, showing a flat and homogeneous Se-terminated FeSe surface. Scale bar, 2 nm. ($V_{\mathrm{b}} = -30$ mV, $I_{\mathrm{t}} = 5$ nA.) **c,** Differential-conductance spectra, $dI/dV$, acquired with the superconducting Nb tip at progressively increasing junction transparency, $1/R_N$ (colour scale). With increasing junction transparency, a pronounced zero-bias Josephson peak, $g_0$, develops, accompanied by negative differential conductance, $g_{dip}$, and finite-bias Andreev conductance, $g_{\mathrm{A}}$, near the Nb superconducting gap. Inset, topography showing the location at which the spectra were acquired (red dot). **d,** Zero-bias Josephson conductance $g_0$ (orange) and finite-bias Andreev conductance $g_{\mathrm{A}}$ (cyan-blue), extracted from **c**, plotted as a function of $1/R_N^2$. Open circles denote experimental data and solid lines are linear fits. The linear dependence of $g_0$ on $1/R_N^2$ is consistent with the Josephson tunnelling, for which $g_0 \propto I_J^2$ and $I_J \propto 1/R_N$, whereas the linear dependence of $g_{\mathrm{A}}$ on $1/R_N^2$ is consistent with two-electron Andreev tunnelling, for which $g_{\mathrm{A}} \propto 1/R_N^2$.

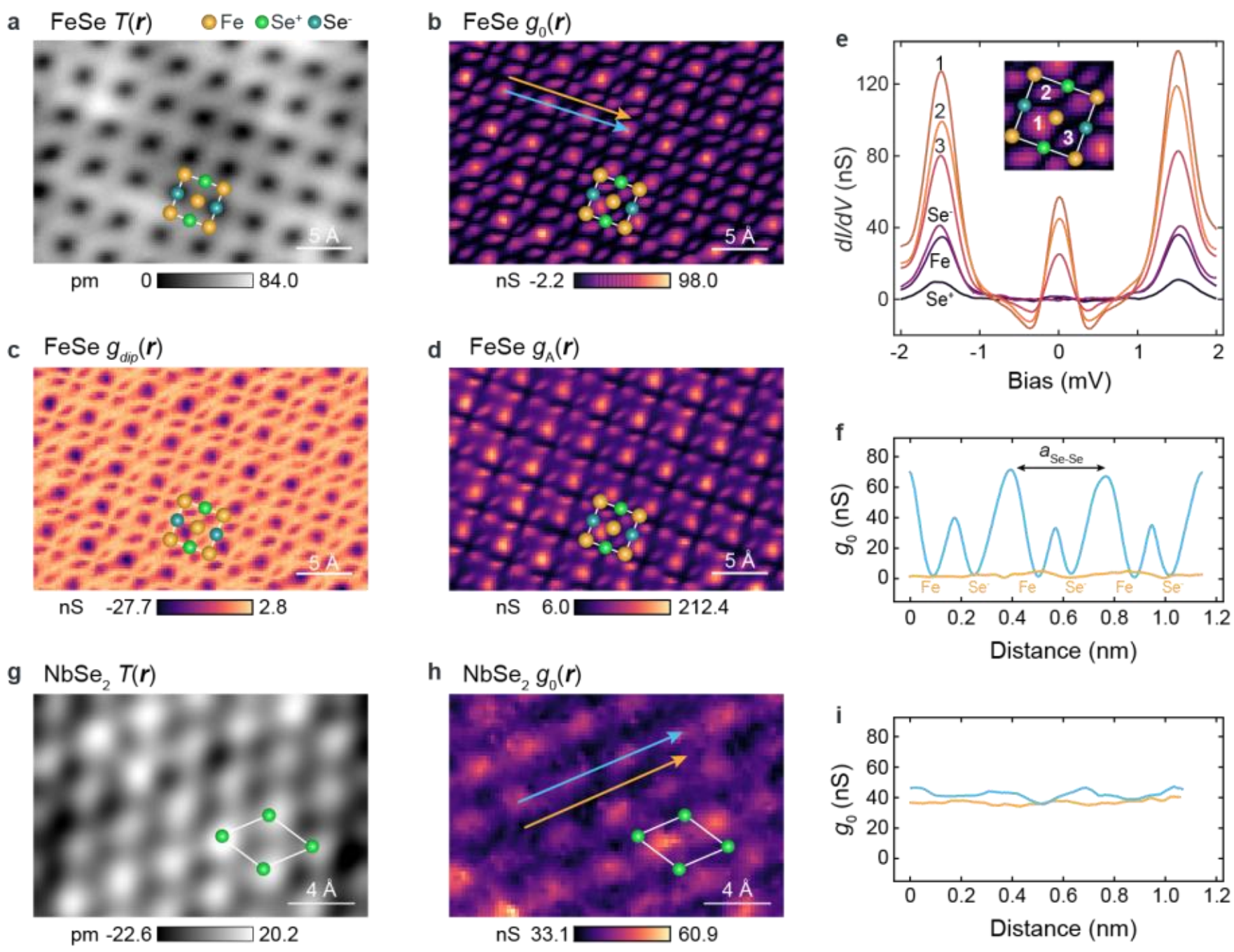

**Fig. 2 | Intra-unit-cell structure of the Josephson response in monolayer FeSe. a–d**, Simultaneously acquired topography, $T(\boldsymbol{r})$ (**a**), zero-bias Josephson conductance map, $g_0(\boldsymbol{r})$ (**b**), conductance map at the negative-differential-conductance dip (0.4 meV), $g_{dip}(\boldsymbol{r})$(**c**), and finite-bias Andreev conductance map, $g_{\mathrm{A}}(\boldsymbol{r})$(**d**), measured over the same region of monolayer FeSe at a fixed junction setpoint of $V_{\mathrm{b}} = -30$ mV and $I_{\mathrm{t}} = 80$ nA, corresponding to normal-state conductance $1/R_N \sim 2.67$ μS. The Josephson response develops nodal lines along the Fe–Se directions, while the Andreev response is strongly suppressed along the corresponding directions. The overlaid structural model identifies the Fe sites and the upper and lower Se sublattices, $Se^+$ and $Se^-$, respectively; the same crystallographic registry is used in all four panels. Scale bars, 5 Å. **e**, Representative *dI/dV* spectra acquired at selected positions within the FeSe unit cell, as indicated in the inset. Spectra 1–3 are taken at three indicated Fe–Fe bond centres, together with spectra acquired at Fe, $Se^+$ and $Se^-$ sites. **f,** Line profiles of $g_0(\boldsymbol{r})$ taken through the Fe–Fe bond centres (cyan-blue) and following the nodal line through Fe and Se sites (orange) along the parallel arrows marked in **b**. Fe and Se positions are marked to register the nodal structure to the lattice. **g**,**h**, Simultaneously acquired topography, $T(\boldsymbol{r})$ (**g**), and zero-bias Josephson $g_0(\boldsymbol{r})$ map (**h**), of $NbSe_2$, measured with the same superconducting-tip configuration at a fixed junction setpoint of $V_{\mathrm{b}} = -5$ mV and $I_{\mathrm{t}} = 10$ nA. The overlaid markers indicate the surface Se lattice and are registered identically in the two panels. Scale bars, 4 Å. **i**, Line profiles of $g_0(\boldsymbol{r})$ along the two lines indicated in **h**, showing only weak atomic-scale variation and no corresponding nodal structure.

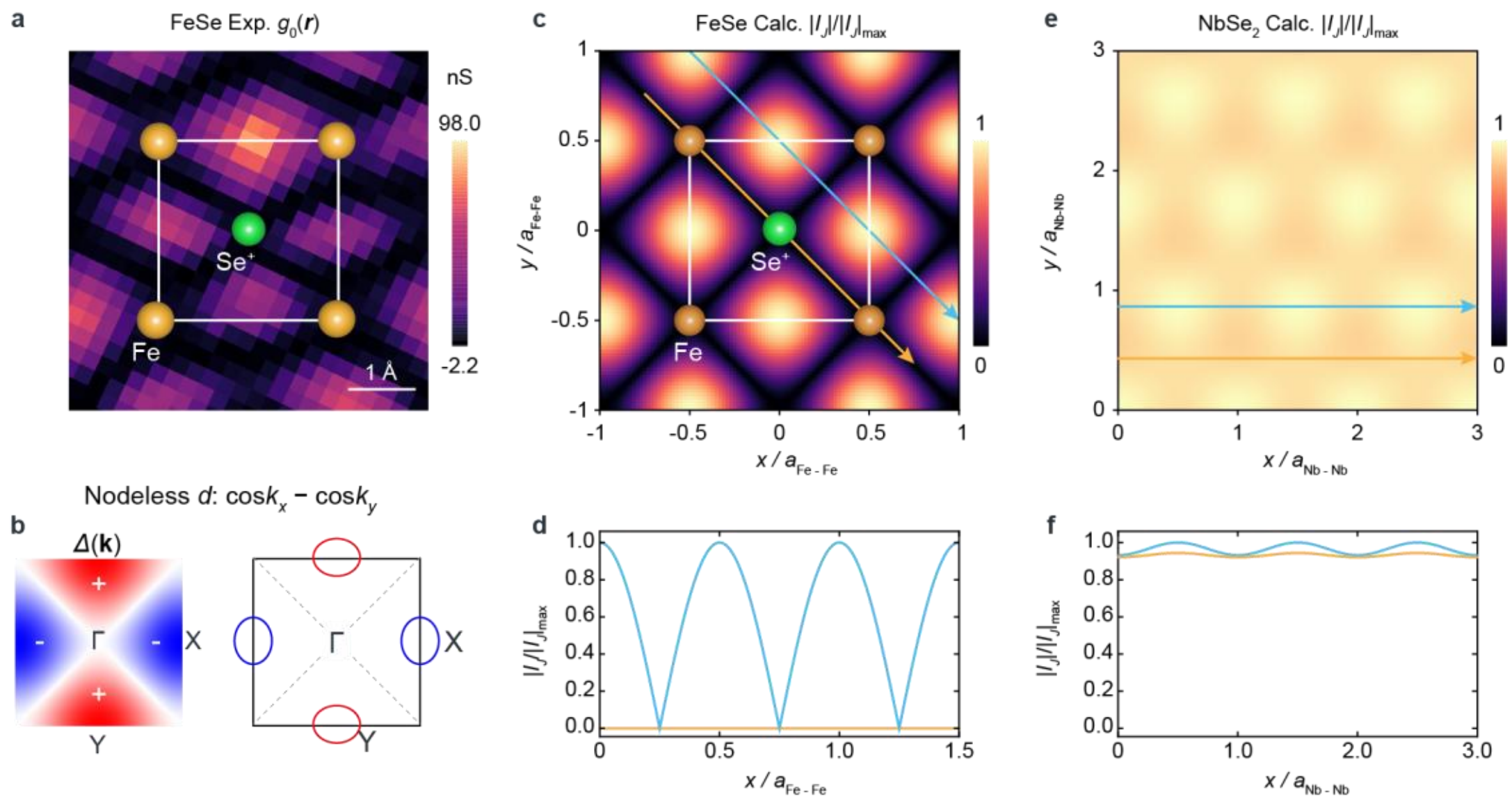

**Fig. 3 | Atomic-scale Josephson interference reveals the internal pairing structure of monolayer FeSe. a,** Experimental Josephson $g_0(\boldsymbol{r})$ map of monolayer FeSe. The overlaid Fe lattice defines the crystallographic registry in the 1-Fe unit-cell description. Scale bar, 1 Å. **b,** Nodeless *d*-wave pairing state used in the microscopic calculation. Left, momentum-space pairing form factor, $\Delta(\boldsymbol{k}) \sim \cos k_x - \cos k_y$, with colour indicating the sign of the order parameter. Right, schematic electron pockets of monolayer FeSe at the X (±π, 0) and Y (0, ±π) points of the unfolded 1-Fe BZ. The nodal lines of $\Delta(\boldsymbol{k})$ avoid the Fermi surfaces, allowing a fully gapped quasiparticle spectrum despite the sign-changing order parameter. **c,** Calculated spatial distribution of the normalized Josephson coupling amplitude, $|\, I_J(\boldsymbol{r}) \,|/|\, I_J \,|_{\max}$, for the $B_{1g}$ state. The calculation reproduces the experimentally observed intra-unit-cell interference pattern, including maxima near the Fe–Fe bond centres and symmetry-related nodal lines. The same Fe-lattice registry as in **a** is overlaid. **d,** Representative line profiles along the paths marked in **c**, showing periodic suppression of the Josephson coupling to zero at the nodal positions. **e,** Corresponding calculation of $|\, I_J(\boldsymbol{r}) \,|/|\, I_J \,|_{\max}$ for $NbSe_2$, used as a conventional fully gapped reference. **f,** Representative line profiles along the paths marked in **e**, showing the absence of the symmetry-related nodal structure found for FeSe. Together, the agreement between experiment and microscopic calculation identifies the intra-unit-cell Josephson pattern as a real-space interference fingerprint of nodeless *d*-wave pairing in monolayer FeSe.

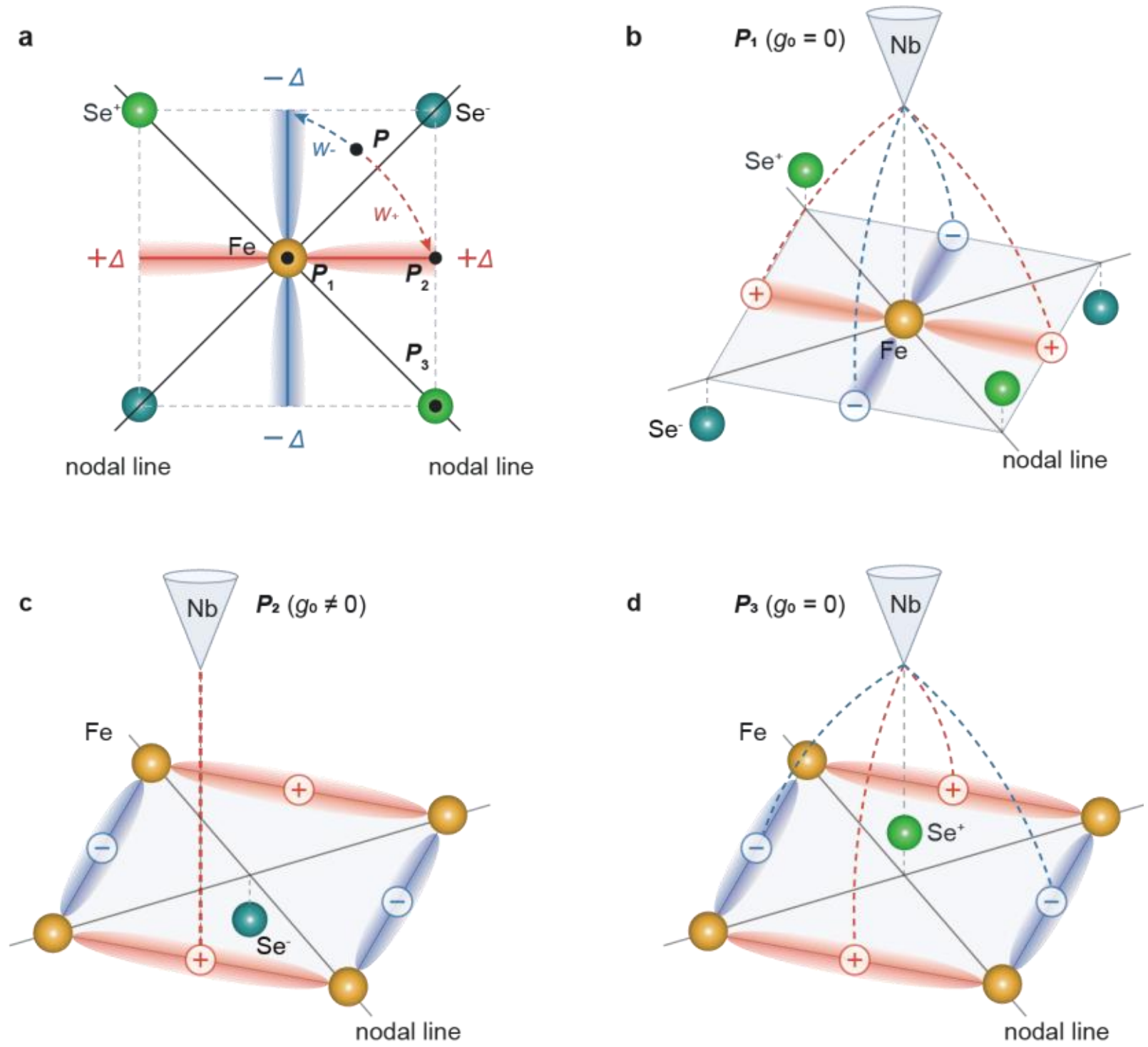

**Fig. 4 | Tip-position-controlled Josephson interference between opposite-sign pairing components. a,** Schematic top view of the nearest-neighbour bond-centred pairing components ($\Delta$) in monolayer FeSe. Red and blue denote pairing amplitudes with relative phases 0 ($+\Delta$) and π ($-\Delta$), respectively, corresponding to opposite signs of the local pair field. The black lines mark the symmetry-related nodal directions along which the net local Josephson coupling vanishes. $w_+$ and $w_-$ represent the tunnelling weights from the tip position ***P*** to the neighbouring bond-centred pairing components. **b–d,** Real-space schematics of the local Josephson coupling between an $s$-wave Nb tip and the opposite-sign nearest-neighbour pairing components $\pm\Delta$ of monolayer FeSe for different tip positions within the unit cell. The red and blue shaded lobes indicate positive- and negative-phase pairing components, respectively, and their relative overlap with the tip determines the local Josephson response. **b,** With the tip above an Fe site ($\boldsymbol{P}_1$), symmetry-related opposite-sign contributions couple with equal weight and interfere destructively, yielding $g_0 = 0$. **c,** With the tip above a nearest-neighbour Fe–Fe bond centre ($\boldsymbol{P}_2$), the tunnelling weights become unequal, lifting the cancellation between opposite-sign contributions and producing a finite Josephson response, $g_0 \neq 0$. **d,** With the tip above a Se site ($\boldsymbol{P}_3$), symmetry-related contributions again couple with equal weight in the idealized geometry, restoring destructive interference and $g_0 = 0$. The tip-position-dependent redistribution of tunnelling weights therefore converts the relative phase of the microscopic pairing components into the observed intra-unit-cell Josephson interference pattern.

## Methods

**Sample preparation.** Monolayer FeSe films were grown on Nb-doped STO(001) substrates by molecular beam epitaxy under ultrahigh-vacuum conditions. The STO substrates were first degassed at 600 °C and subsequently annealed at 1200 °C to obtain an atomically flat $TiO_2$-terminated surface. FeSe films were grown by co-evaporating Fe and Se from separate evaporation sources onto the STO substrates held at ~400 °C. A Se-rich growth condition was used, with the Se flux maintained substantially higher than the Fe flux. The deposition rate was calibrated such that a growth time of approximately 10 min corresponded to one unit-cell layer of FeSe. After growth, the films were annealed in vacuum at approximately 500 °C for ten hours before STM measurements.

**STM measurements.** STM/S measurements were performed using a dilution-refrigerator STM operating at a base temperature of approximately 30 mK under ultrahigh-vacuum conditions with a base pressure below $5 \times 10^{-11}$ Torr. Superconducting Nb tips were prepared *in situ*. Before STM measurements, Nb tips were sharpened by electron-beam heating in the molecular-beam-epitaxy chamber and subsequently conditioned on a clean Nb single-crystal surface. The Nb(100) surface was prepared by repeated cycles of $Ar^{+}$ sputtering and annealing under ultrahigh vacuum. Tip conditioning was performed by controlled indentation and voltage-pulse procedures on the clean Nb surface until a stable superconducting tunnelling spectrum was obtained. The superconducting character of the tip was verified by tunnelling spectroscopy on the Nb single-crystal surface. The resulting superconductor–insulator–superconductor spectrum exhibited well-defined coherence peaks associated with the combined superconducting gaps of the Nb tip and Nb sample, confirming a superconducting Nb tip with high spectroscopic quality.

Topographic images were acquired in constant-current mode at the specified sample bias $V_\mathrm{b}$ and tunnelling current $I_\mathrm{t}$. $dI/dV$ spectra and maps were measured using a standard lock-in technique by adding a small a.c. modulation $V_\mathrm{mod}$ at a frequency of 973 Hz to the d.c. sample bias. For spatially resolved Josephson measurements, the tip–sample distance was stabilized at the specified junction resistance $R_N = |V_\mathrm{b}/I_\mathrm{t}|$ before opening the feedback loop, and the zero-bias differential conductance, $g_0$, was recorded as the local Josephson response.

**Calculations.** The spatial Josephson response of monolayer FeSe was calculated using a minimal single-orbital model[40,50]. We neglected spin–orbit coupling and used the unfolded one-Fe convention, with the $x$ and $y$ axes along nearest-neighbour Fe–Fe bonds and the Fe–Fe spacing set to unity. The normal-state Hamiltonian is

$$h_0(k) = \left[2t_1\left(\cos k_x + \cos k_y\right) + 4t_2 \cos k_x \cos k_y - \mu\right] s_0,$$

where $t_1$ and $t_2$ denote nearest- and next-nearest-neighbour Fe–Fe hopping, respectively, $\mu$ is the chemical potential, and $s_0$ and $s_i$ are the identity and Pauli matrices in spin space. Superconductivity was introduced through the $d$-wave spin-singlet pair potential

$$\Delta_d(\boldsymbol{k}) = 2\Delta(\cos k_x - \cos k_y) i s_2,$$

where $\Delta$ is the nearest-neighbour bond pairing amplitude.

We calculated the lattice anomalous Green's function from the Bogoliubov–de Gennes (BdG) Hamiltonian and obtained the continuum propagator at the tip position $\boldsymbol{r}$ through the Wannier expansion[15],

$$F_s(\boldsymbol{r}, \boldsymbol{r}, i\omega_n) = \sum_{R,R'} F_s(R, R', i\omega_n)\, w_R(\boldsymbol{r}) w_{R'}(\boldsymbol{r}),$$

where $\omega_n = (2n+1)\pi\, k_B T$ are the fermionic Matsubara frequencies at temperature $T$. The lattice propagator $F_s(R, R', i\omega_n)$ describes the microscopic pair correlations between sites $R$ and $R'$, and $w_R(\boldsymbol{r})$ is the Wannier function centred at $R$, evaluated at the tip position. The Wannier functions provide spatial tunnelling weights, while the coherent sum preserves the relative phases of the pair correlations. The local Josephson coupling to the $s$-wave Nb tip was then calculated from[15]

$$I_J(\boldsymbol{r}) \propto k_B T \sum_{n=-\infty}^{\infty} \mathrm{tr}_{\mathrm{spin}}\left[F_t^\dagger(i\omega_n) F_s(\boldsymbol{r}, \boldsymbol{r}, i\omega_n)\right],$$

where $F_t^\dagger$ is the tip pair-creation anomalous Green's function and $\mathrm{tr}_{\mathrm{spin}}$ denotes the trace over spin. The tip was modelled with an isotropic BCS gap and a constant normal-state density of states over a finite bandwidth.

For $NbSe_2$, we used an effective bilayer tight-binding model with three Nb $d$ orbitals per layer, interlayer hopping and Ising spin–orbit coupling[51]. Its parameters were fitted to published data[52,53]. A uniform, orbital-diagonal onsite spin-singlet $s$-wave pair potential was introduced in both layers. We calculated the Josephson response using the same tunnelling formalism, with orbital-dependent weights coupling the tip to the upper layer and a coherent sum over sites and orbitals.

The calculated maps and line profiles show $|I_J(\boldsymbol{r})|/|I_J|_{\max}$, normalized separately for each material; the measured zero-bias conductance satisfies $g_0(\boldsymbol{r}) \propto |I_J(\boldsymbol{r})|^2$. Model parameters, tunnelling weights and Green's-function conventions are provided in the Supplementary Information (Section S1–S4 and Tables 1 and 2).

**Data availability**

All data are available in the main text and Supplementary Information, and from the corresponding authors upon reasonable request.

**Method references**

**Acknowledgements**

This work was supported by the National Key Research and Development Program of China (2024YFA1408102, 2022YFA1403101), Guangdong Provincial Quantum Science Strategic Initiative (GDZX2502001), National Natural Science Foundation of China (12304164, 12674164), the Shenzhen Natural Science Foundation (JCYJ20240813094203005), the New Cornerstone Investigator Program, and the Chinese Academy of Sciences Project for Young Scientists in Basic Research (2022YSBR-048).

**Author contributions**

M.-Q.R., Y.-J.S. and Q.-K.X. conceived the project. R.L., Y.H., M.W. and M.-Q.R. conducted the STM experiments under the supervision of Y.-J.S. and M.-Q.R.. J.M. and K.J. performed the theoretical calculations and simulations. R.L., Y.H. and M.-Q.R. performed the data analysis. M.-Q.R., K.J. and Y.-J.S. wrote the manuscript with input from Q.-K.X.. All authors contributed to the discussion of the results and to revising the manuscript. Q.-K.X. and Y.-J.S. supervised the project.

**Competing interests**

The authors declare no competing interests.

Correspondence and requests for materials should be addressed to M.-Q.R., K.J. and Y.-J. S.